\documentclass[12pt, letterpaper]{article}

\usepackage{amsmath,amssymb}
\usepackage{cite}
\usepackage{fancyhdr}
\usepackage[top=1in, bottom=1in, left=1in, right=1in]{geometry}
\usepackage{graphicx}
\usepackage{hyperref}

\numberwithin{equation}{section}
\date{}

\begin{document}
\title{{\rm\footnotesize \qquad \qquad \qquad \qquad \qquad \ \qquad \qquad \qquad \ \ \ \ \ \                      RUNHETC-2024-31
}\vskip.5in    Schwinger-Dyson Resummation of Perturbation Expansions}
\author{Tom Banks\\
NHETC and Department of Physics \\
Rutgers University, Piscataway, NJ 08854-8019\\
E-mail: \href{mailto:tibanks@ucsc.edu}{tibanks@ucsc.edu}
\\
\\
}

\maketitle
\thispagestyle{fancy} 

\begin{abstract} \normalsize \noindent  We introduce a hierarchical system of approximations for resumming both conventional semiclassical and large $N$ vector expansions of models in quantum field theory and condensed matter physics.  Each stage of the hierarchy consists of a closed set of non-linear equations for one particle irreducible correlation functions with no more than $K$ points and captures the perturbative expansion of each of those functions up to some finite order.  As $K \rightarrow \infty$ the full Schwinger-Dyson hierarchy is approached. We present an argument that for ordinary finite dimensional integrals, the procedure converges to the exact answer in this limit, despite the fact that no finite order of the hierarchy sees "instanton" effects.  Some potential applications to the calculation of critical exponents, to low dimensional condensed matter systems, and to the phase structure of the homogeneous electron gas are outlined, but no detailed calculations are presented. \end{abstract}


\newpage
\tableofcontents
\vspace{1cm}

\vfill\eject
\section{Introduction}

The familiar functional integral formula for correlation functions in quantum field theory is the solution of the Schwinger-Dyson (SD) equations for those functions.  Different field theory states correspond to different functional integration contours.  Semi-classical and vector $1/N$ expansions are generated by steepest descent evaluation of the functional integral.  When the SD equations for one particle irreducible correlators are expanded around the stationary point, one sees that the $k$ point functions of fluctuations begin at order $k$ in the expansion parameter.  

The exact SD equations always form an infinite hierarchy, relating the $k$ point function to functions with $k+p$ points up to at least some finite $p$.  Given an expansion scheme however, we can define a systematic set of truncations of these equations, by neglecting functions with more than $K$ points, with some $K$.   Each of these truncations gives a closed set of non-linear equations for $k \leq K$ point functions, which sum up an infinite number of Feynman diagrams of the original model.  

As a simple example, let's consider the integral
\begin{equation} Z[j] = \int\ dx e^{ - \frac{m^2}{2} x^2 - \frac{g^2}{4!} x^4 + j x} . \end{equation}
This defines a function of $g^2$ that is positive on the real axis and analytic in the plane cut along the negative real axis.  
Correlation functions $Z_n$ are defined as $n$th derivatives of $Z$, divided by $Z[0]$.
$Z[j]$ satisfies
\begin{equation} (m^2 \partial_j + \frac{g^2}{3!} \partial_j^3) Z = j Z . \end{equation}
If $Z \equiv e^W$, then
\begin{equation} (m^2 \partial_j W + \frac{g^2}{3!}[ (\partial_j W)^3 + 3 (\partial_j W) \partial_j^2 W) + \partial_j^3 W] = j . \end{equation}  Taking $2n-1$ derivatives of this and setting $j = 0$ we get the SD equations for connected correlation functions. 

These equations have an exact scaling symmetry under $m^2 \rightarrow a m^2$, $g^2 \rightarrow b g^2$, $j \rightarrow c j$, with $a = \frac{b}{c^2} = c^2$.  This implies that 
\begin{equation} W[m^2, g^2, j] = W[1,\frac{g^2}{m^4}, \frac{j}{m}] . \end{equation}  The $2n$ point correlator thus behaves like $W_n = m^{-2n}  f_n (\frac{g^2}{m^4})$.  On the other hand, it's clear from the original integral that the limit of $W_n$ as $m^2 \rightarrow 0$ is finite, so we know the large $g^2$ behavior of the functions $f_n$ up to a constant multiple.

The Legendre transform of $W$,
\begin{equation} W[j] = G[f] + f j , \end{equation} satisfies
\begin{equation} m^2 f +  \frac{g^2}{3!}[f^3 - 3f G_2^{-1} - G_2^{-3} G_3] = - G_1 , \end{equation} where
\begin{equation} G_k \equiv \partial_f^k G . \end{equation}  The first non-trivial equation for correlation functions comes from taking one derivative of this and setting $f = 0$ 
\begin{equation} - G_2 = m^2 - \frac{g^2}{3!}[3 G_2^{-1} + G_2^{-3} G_4] . \end{equation} 
If we take two more derivatives and set $f = 0$ we get an equation for $G_4$, 
\begin{equation} G_4 = - \frac{g^2}{6}[ 6 + 9 G_2^{-4} G_4^2 + 9 G_2^{-2} G_4 - G_2^{-1} G_6].\end{equation}  Since the equation for $G_6$ implies that it's perturbation expansion starts at order $g^6$, we can truncate the equations by setting it equal to zero in the equation for $G_4$.  We then get a closed set of non-linear equations for $G_2$ and $G_4$.  Writing
$G_2 = g Y_2 $, $G_4 = g^2 Y_4$ we find

\begin{equation} 3 Y_4^2 + (3 + 2 Y_2) Y_2 Y_4 + 2 Y_2^2 = 0 . \end{equation}
\begin{equation} g Y_2  = - m^2 + \frac{g}{6}[Y_2^{-3} Y_4 + 3 Y_2^{-1}] . \end{equation}
An interesting feature of these equations is that they preserve the scaling symmetry of the exact equations and so have the correct qualitative behavior in the $m^2 \rightarrow 0 $ limit.  Of course, the constant value of the coefficient of $Y_{2,4}$ in the limit will not be exactly correct.  However, every subsequent truncation of the equations, in which we neglect only $G_{2K}$ for large enough $K$, will preserve the qualitative strong coupling behavior.  Thus we may expect the constant value obtained by successive truncations of the hierarchy to converge fairly rapidly to the correct answer.

Indeed, there is a non-rigourous argument that this truncation procedure should be convergent for computing $G_{2k}$ with fixed $k$.  These functions are analytic functions of $g^2$ in the cut plane.  We can imagine that the truncations are approximating the equations for them expanded around a small but non-zero value of $g^2$ .   That perturbation series should be convergent, and for any fixed $k$, as we increase the value of $G_{2K}$ that we set to zero in the $K$ th truncation scheme, we are including more and more of the terms in the perturbative expansion in our approximate resummations.  

\section{General Form of the Leading Order Equations}

We will write the equations for a single scalar field $\phi (i)$, which is the fluctuation around the saddle point. This could be an elementary scalar, or the Hubbard-Stratonovich scalar of a large $N$ vector model.  Similarly the coupling $g$ that appears in the action below could be a semi-classical expansion parameter or $N^{-\frac{1}{2}}$.
$i$ represents a space-time lattice point.  Partial derivatives w.r.t. $\phi (i)$ become functional derivatives in a continuum field theory, sums over $i$ (which are generally left implicit by the Einstein convention) become integrals, and multi-index objects become functions of multiple space-time variables
The imaginary time action for $\phi$ is
\begin{equation} S[\phi, J] = \sum_{k=2}^{\infty} \frac{g^{k - 2}}{k!} S^{(k)} (i_1 \ldots i_k) \phi (i_1) \ldots \phi (i_k)  + J(i) \phi (i) . \end{equation}  The generating functional is defined by
\begin{equation} Z[J] \equiv\frac{\int [d\phi (i)] e^{- S[\phi, J]} }{\int [d\phi (i)] e^{- S[\phi, 0]} }. \end{equation}
The SD equations are derived by setting the integral of $\partial / \partial \phi (i) $ of the numerator integrand equal to zero, and then rewriting functions of $\phi (i)$ as derivatives of $Z$ w.r.t. $J(i)$.  To leading order in $g$ we get the following closed set of equations for the one and two point functions in a general model, and for the two and four point functions in a model with a $\phi \rightarrow - \phi$ symmetry.  

\begin{equation}  J(i) + Z^{-1}[ S^{(2)} (i,j) \frac{\partial Z}{\partial J (j)} + \frac{g}{2} S^{(3)} (i,j,k) \frac{\partial^2 Z}{\partial J (j) \partial J (k)}] = 0 . \end{equation}
\begin{equation}  J(i) + Z^{-1}[ S^{(2)} (i,j) \frac{\partial Z}{\partial J (j)} + \frac{g^2}{3!} S^{(4)} (i,j,k,l) \frac{\partial^2 Z}{\partial J (j) \partial J (k) J (l)}] = 0 . \end{equation}
If $\phi$ is an elementary field, $S^{(2)}$ is just the usual Klein-Gordon operator, while for a Hubbard Stratonovich field it will be a bare kinetic term plus a one loop "RPA" contribution.  
To write these in a more useful form, we make the usual definitions
\begin{equation} Z[J] = e^{- W[J]} = e^{- [G[f] + J(i) f(i)]} . \end{equation}
\begin{equation} f(i) = \frac{\partial W}{\partial J(i)}, \ \ \ \frac{\partial G}{\partial f(i)} = - J(i) . \end{equation}
We then rewrite the equations as equations for $G$ in terms of $f(i)$, take one more derivative w.r.t. $f(i)$ to get a second equation, set $J(i) = 0$ and keep only leading order in $g$.   In the models with no reflection symmetry this gives
\begin{equation} S^{(2)} (i,j) f_0 (j) = \frac{g}{2} S^{(3)} (i,j,k)[f_0 (j) f_0(k) -  \frac{\partial^2 W}{\partial \phi (j) \partial \phi (k)}] . \end{equation}
\begin{equation} G^{(2)} (i,j) = S^{(2)} (i,j) + g S^{(3)}(i,j,k) f_0 (k) + \frac{g^2}{2} S^{(3)} (i,m,k) G^{(2)\ - 1} (k,a) G^{(2)\ -1} (b,m) S^{(3)} (a,b,j) . \end{equation}  $f_0(i)$ is the tadpole of the field $\phi (i)$, the fluctuation correction to the saddle point value $\Phi$.  In the second term we've approximated $G^{(3)}$ by $g S^{(3)}$, its leading order contribution.  The next step in the systematic hierarchy, would keep the leading order contribution $G^{(4)} = g^2 S^{(4)}$ and treat $G^{(3)}$ as an unknown function.  
As an aside, we note that in\cite{tblargeNresum} a somewhat different hierarchy of partial resummations of large $N$ expansions was proposed.  At present we do not understand the relation between the two approaches, but the present procedure is easier to implement because it does not require one to integrate the solution of a non-linear integral equation over values of the 't Hooft coupling.  

For models with a reflection symmetry, the leading order approximation sets $G^{(4)} = g^2 S^{(4)}$ and we get a closed equation for the two point function
\begin{equation} G^{(2)} (i,j) = S^{(2)} (i,j) + \frac{g^2}{2} S^{(4)} (i,j,a,p) G^{(2)\ - 1} (p,a) \end{equation}\begin{equation} + \frac{g^4}{3!} S^{(4)} (i,m,n,p) G^{(2)\ - 1} (p,a) G^{(2)\ - 1} (n,b)G^{(2)\ - 1} (m,c) S^{(4)} (a,b,c,j) . \end{equation} 
However, there is often a phase of the system where the reflection symmetry is spontaneously broken.  In this phase there is a more complicated truncation of the SD equations, which we will deal with in the example below.

\section{Application to Scalar Field Theory}

Consider a scalar field $\Phi$ in $D$ Euclidean dimensions, with action
\begin{equation} S = \int d^D x\ [ \frac{1}{2} (\nabla \Phi)^2  + m_0^2 \Phi^2) + \frac{g^2}{4!} \Phi^4 ] . \end{equation}  If $m_0^2 > 0$ we just get the equation for the two point function
\begin{equation} G^{(2)} (x - y) = (- \nabla^2 + m_0^2) \delta^D (x - y) + \frac{g^2}{2} G^{(2)\ - 1} (0)  \delta^D (x - y) + \frac{g^4}{3!} [G^{(2)\ - 1} (x - y)]^3 . \end{equation} 
For $m_0^2 < 0$ we expand around the stationary point $\Phi_0 = \sqrt{ - 3! m_0^2 / g^2}$.  The stable excitation has a mass $\mu^2 = -  m_0^2 /2$ and a cubic coupling $\frac{g|m_0|}{\sqrt{3!}} $.  There are also non-trivial tadpole corrections to the expectation value $\Phi_0$.  The leading order equations for the one and two point functions are
\begin{equation} \frac{m_0^2}{2}\phi +\sqrt{3/2} g |m_0| \phi^2 = \sqrt{3/2}|m_0| g G^{2\ - 1} (0) . \end{equation}   
\begin{equation} G^{(2)} (x - y) = (-\nabla^2 + \frac{m_0^2}{2}) \delta^D (x - y) + \sqrt{6} |m_0| \phi \delta^D (x - y) - \sqrt{3/2} g^2 m_0^2 [G^{2\ - 1} (x - y)]^2 + \frac{g^2}{2} G^{(2)\ - 1} (0)  \delta^D (x - y) \end{equation} \begin{equation} + \frac{g^4}{3!} [G^{(2)\ - 1} (x - y)]^3 
. \end{equation}   The inverse of $G^{(2)}$ in these equations is the kernel of its inverse as an integral operator.  

These equations contain obvious UV divergences coming from the factor $G^{2\ - 1} (0)$ and from the powers of $G^{2\ - 1} (x - y)$, which when $g = 0$, is a singular distribution rather than a function.  We imagine these are cut off by some regularization scheme.  The most interesting question is what happens near the transition point where one has to switch between one perturbation scheme and the other.  The standard lore tells us that this transition actually occurs at non-zero values of $m_0^2$ and $g^2$ when $D < 4$, but for fixed cut off, they are close to this point if we allow $D$ to be a continuous variable near $4$.  

Thus, we might expect that the equations we have written provide a non-perturbative resummation of the $\epsilon$ expansion for critical exponents of the magnetization and the two point correlator of spin fluctuations of models in the Ising universality class.  For integer $D < 4$ the above equations are rendered finite by the usual UV renormalization procedure, which only involves a finite number of graphs.  Note however that the loop corrections to $G^{(4)}$ are not included at this level of truncation.  Thus, in order to capture the full renormalization group flow and find the non-trivial infrared fixed point, it is likely that we have to go to the next order approximation.

\section{Hertz-Millis Models}

Hertz-Millis models, in which a critical boson $\phi$ couples to a fermi surface in two spacial dimensions, have long been considered interesting candidates for quantum critical points that might underlie strange metal behavior and other exotic experimental phenomena in condensed matter physics.  They have resisted traditional field theoretic approaches to the solution of their infrared behavior for more than twenty years.  In particular\cite{sslee}, showed that the standard vector large $N$ expansion has large corrections in the IR, which have so far resisted attempts at resummation.  

We will introduce a new large $N$ technique, which might have better behavior.  We define a pair of bilocal fields $\lambda (t,\vec{x},t^{\prime} \vec{x^{\prime}}), \chi (t,\vec{x},t^{\prime} \vec{x^{\prime}})$.   The expectation value of $\chi$ is just 
\begin{equation} N^{-1} \langle 0 | T \psi_a (x,t) \psi_a^{\dagger} (x^{\prime}, t^{\prime}) | 0 \rangle . \end{equation} Expanding around a constant self consistent solution $\phi_0$ for the Hertz-Millis field, we find that we can eliminate the fluctuations of both $\chi$ and $\lambda$ and compute correlation functions of fluctuations of $\chi$ efficiently in terms of an effective action for the local boson.  The non-quadratic terms in this action are suppressed by powers of $N^{-\frac{1}{2}}$.  This allows us to write a systematic $1/N$ expansion for correlators of $\chi$ as well as closed equations (the first of a systematic infinite hierarchy of closed equations)  for the one and two point functions of fluctuations of $\chi$, which re-sum an infinite number of terms in their large $N$ expansions.  

\subsection{The Effective Action}

After introducing auxiliary fields and integrating out the fermions in the Hertz-Millis model we obtain an imaginary time action
\begin{equation} S = N [{\rm Tr\ ln\ } (\partial_t - \epsilon - g \hat{\phi} - \lambda) - \frac{1}{2} \int\ dt d^2 x (\phi (\square + m^2) \phi) + \int\ dt\ dt^{\prime}\ d^2 x\ d^2{x^{\prime}} \lambda (t,\vec{x}; t^{\prime}, \vec{x^{\prime}}) \chi ( t^{\prime}, \vec{x^{\prime}} ; t,\vec{x})] . \end{equation} 
$\epsilon $ is an angle dependent linear function of spatial derivatives, which defines the Fermi surface.  $g$ is imaginary, since $\phi$ was coupled to a charge density.  The generalization to fermion fields with multiple spin components and critical bosons coupled to other types of bilinears is straightforward.  $\hat{\phi}$ is the integral operator
\begin{equation} \hat{\phi} ( t^{\prime}, \vec{x^{\prime}} ; t,\vec{x}) \equiv \phi (t,\vec{x}) \delta (t - t^{\prime}) \delta^2 (x - x^{\prime}) . \end{equation}  We will often simplify notation by writing integrals and delta functions in three dimensional space-time. 

We look for translation invariant stationary points for the variables and find
\begin{equation} \lambda_0 = 0, \ \ \ \ g\phi_0 = \langle t, \vec{x} | (\partial_t - \epsilon - g\hat{\phi_0})^{-1} | t, \vec{x} \rangle . \end{equation} 
\begin{equation} \chi_0 = \langle t, \vec{x} | (\partial_t - \epsilon - \hat{\phi_0})^{-1} | t^{\prime}, \vec{x^{\prime}} \rangle . \end{equation}  

Now we write
\begin{equation} \chi = \chi_0 + N^{- \frac{1}{2}} \Delta_{\chi}, \end{equation}
\begin{equation} \phi = \phi_0 + N^{- \frac{1}{2}} a, \end{equation}
\begin{equation} \lambda = \hat{\phi_0} + N^{- \frac{1}{2}} \Delta_{\lambda}. \end{equation}
The effective action for fluctuations, including a source for $\Delta_{\chi}$, is
\begin{equation} S = N {\rm Tr\ ln\ } (\partial_t - \epsilon - g\hat{\phi}_0 + (\Delta_{\lambda} + g\hat{a}) N^{- \frac{1}{2}} ) - N^{\frac{1}{2}} {\rm Tr\ }[\chi_0 (\Delta_{\lambda} + g\hat{a})] + \frac{1}{2} \int a (\square + m^2) a \ \ \end{equation}
\begin{equation}+ \rm{Tr}\ [ (\Delta_{\lambda} + J_{\chi})\Delta_{\chi})]  . \end{equation}  We can now do the functional integral over $\Delta_{\chi}$ which sets $(\Delta_{\lambda} + J_{\chi}) = 0$.  We're left with an integral over the local field $a$ coupled to the bilocal source $J_{\chi}$ for fluctuations of $\chi - \chi_0$.  We can also add a local source $J$ for $a$.   The imaginary time action is then,
\begin{equation} \int \frac{1}{2} [a(\square + m^2) a  + a J] + N {\rm Tr\ } [{\rm ln\ } (\partial_t - \epsilon - g\hat{\phi}_0 + N^{-\frac{1}{2}} (g\hat{a} - J_{\chi})) - {\rm Tr\ } [{\rm ln\ }  N^{\frac{1}{2}}\chi_0 (g\hat{a} - J_{\chi})] .\end{equation}
Note that in the terms coming from integrating out the fermions, a shift in the integration variable $a(x)\rightarrow a(x) + \delta a(x) $ can be compensated by a shift in the bi-local source term $J_{\chi} \rightarrow J_{\chi} +g \delta a(x) \delta^3 (x - y) $.  Thus, we obtain the Schwinger Dyson equation for the generating functional
\begin{equation} (\square + m^2)\frac{\delta}{\delta J(x)} Z [J,J_{\chi}] = J(x) Z [J,J_{\chi}] + \frac{\delta}{g \delta J_{\chi} (x,x)} Z [J,J_{\chi}] . \end{equation}

We obtain a second equation by noting the identity
\begin{equation} \frac{\delta}{\delta J_{\chi} (x,x^{\prime})} {\rm Tr\ } [{\rm ln\ } (\partial_t - \epsilon + g\hat{\phi}_0 + N^{-\frac{1}{2}}(g\hat{a} - J_{\chi})) - N^{-\frac{1}{2}}{\rm Tr\ } [\chi_0 (g\hat{a} - J_{\chi})]  =  \end{equation}
\begin{equation} - N^{-\frac{1}{2}} (\partial_t - \epsilon + g\hat{\phi}_0 + N^{-\frac{1}{2}}(g\hat{a} - J_{\chi}))^{-1} (x,x^{\prime}) + N^{-\frac{1}{2}} \chi_0 (x- x^{\prime}) .\end{equation}  This leads to a second Schwinger-Dyson equation
\begin{equation}  \frac{\delta Z}{\delta J_{\chi} (x,y)} = N^{\frac{1}{2}} \langle [\partial_t - \epsilon - \hat{\phi}_0 - N^{-\frac{1}{2}}(\hat{a} - J_{\chi})]^{-1}  \rangle - N^{\frac{1}{2}} \chi_0 (x - y) .\end{equation} The angle brackets in this equation mean functional average over $a$, with the weight $e^{-S}$.  The space time arguments in the first term are the matrix elements of the inverse differential operator. To any finite order in the $1/N$ expansion these can be written in terms of a finite number of functional derivatives of $Z$, with respect to the local source $J(x)$.
We can use the first equation to convert functional derivatives w.r.t. $J(x)$ into derivatives w.r.t. $J_{\chi} (x,y)$ at equal argument, and then set $J(x) = 0$, obtaining a closed equation for correlation functions of the bilinear fluctuation $\Delta_{\chi}$.  Transport coefficients can all be computed from the two point function of $\Delta_{\chi}$.

Unfortunately, I now think it is unlikely that these equations will lead to a controlled approximation to the IR behavior of Hertz-Millis models.   Consider the one fermion loop diagrams with $k$ vertices.  They have the form
\begin{equation} N^{-k/2 + 1} \int d^2 q\ d \omega \prod_{i = 1}^k \frac{1}{\omega - \omega_i - \vec{\alpha_i} \vec{q - k_i} } . \end{equation}  Here we've assumed that the momenta carried by the boson fields at each vertex is such that all of the internal fermions remain close to the fermi surface.  $\omega_i$ and $k_i$ are linear combinations of those momenta.  Now let's examine a limit in which energies and momenta scale to zero at fixed ratio.  If $a$ is the scaling parameter, the integral behaves like $a^{3 - k}$ , so if $a$ is of order $N^{-1/2}$, then all one loop diagrams look equally important.    Our approximate equation drops all diagrams with $k >3$, and so cannot reproduce the correct IR behavior.

\section{The Homogeneous Electron Gas}

This section is an update and improvement on\cite{tblargeNresum}.  The effective action for the Coulomb field in the homogeneous electron gas, with N flavors of electron is
\begin{equation} S_{eff} = N[- \int \phi \nabla^2 \phi + {\rm Tr\ ln\ } [i\partial_t - \phi - \nabla^2 - \mu]  . \end{equation}  There is a homogeneous positive background charge that cancels off the total charge of the electrons.  It is well known that this model has at least two phases: a weakly coupled fluid phase at high density and the low density Wigner crystal.  Arguments have been presented\cite{colloids}, most persuasively in two spatial dimensions that there is an intermediate colloidal or micro-emulsion phase at intermediate densities.  There is some evidence that the transition between the fluid and the colloid is second order.  Just above the fluid colloid transition, the colloid consists of a solution of stable crystalline fragments with negative surface tension, immersed in a fluid background\cite{ks}\cite{bz3}.  The conjecture is that these fragments become microscopic gapless bosons at the transition.  They should show up in the two point function of the density operator.

The large $N$ expansion of this model is generally used only to show the existence of Debye screening.  In\cite{bz1} it was shown that it cannot exhibit the Wigner crystal phase because it treats the electron density as a classical continuous function.  The Thomas-Fermi equations of the leading large $N$ approximation do not even have meta-stable periodic solutions. 

Indeed, those equations take the form, for static solutions, of the Hartree equations for the electron density functional
\begin{equation} F[n(x)] = F_0 [n(x)] + \frac{e^2}{2} \int dx dy \frac{(n(x) - n_0) (n(y) - n_0)}{|x - y|} , \end{equation} where $F_0$ is the expectation value of the fermion kinetic energy in states constrained to have
\begin{equation} N^{-1} \langle s | \sum_i \psi_i^{\dagger} (x) \psi_i (x) | s \rangle = n(x), \end{equation}
\begin{equation} \int dx (n(x) - n_0) = 0 . \end{equation}
Both functionals on the right hand side are convex, non-negative, and vanish when $n(x) = n_0$.  

The Wigner crystal owes its existence to the discrete pointlike nature of electrons and is missed by the Hartree approximation.  The crystal exists, at large $N$ only when $n_0 \sim N^{-1}$ and the Hartree approximation breaks down.

The leading order equations of our resummation scheme give
\begin{equation} (-\nabla^2 ) f_0 (x) + \int d^D y\ \Pi_2 (x - y) f_0 (y)= \frac{1}{2N^{1/2} }\int d^D y\ d^D z\ \Pi^{(3)} (x,y,z)[f_0 (y) f_0(z) + G^{(2)\ - 1} (y,z)] . \end{equation} 
\begin{equation} G^{(2)} (x,y) = d^{2D} w\ d^{2D} z\ \Pi_3 (x, w_1, w_2) G^{(2)\ - 1} (w_1 - z_1)G^{(2)\ - 1} (w_2 - z_2)\Pi_3 (z_1, z_2, y) .  \end{equation}  Here $f_0 (x)$ is the fluctuation correction to the potential produced by the homogeneous leading order electron charge density.
The $\Pi_k$ are $k$ point one loop fermion diagrams, with propagator $-\nabla^2 + \mu$.  $\mu$ is the chemical potential, and we are interested in $D = 2,3$.  These functions are translation invariant, but the equations also have solutions that only preserve a discrete subgroup of that symmetry.    By assumption, the expectation value $f_0 (x)$ of the potential in these equations is of order $1/N$, and corresponds to an electron density of the same size.  

If we plug either the homogeneous or periodic solutions into the functional $G[f]$, we get an estimate of the lowest energy of a periodic state. 
\begin{equation} E_{fluid/per}  = \frac{1}{2} \int d^D x d^D y\  f_0(x) f_0 (y) G_{fluid/per}^{(2)} (x,y) . \end{equation}
$f_0$ is the appropriate solution for the one point function in the two cases.  
Since the leading order large $N$ equations give exactly zero for the energy of the stationary point solution, this is the leading order large $N$ correction to the energy of the fluid phase, as well as the only evidence at large $N$ for the periodic phase.
At large finite $N$, the periodic state will be meta-stable if $E_{per} > E_{fluid}$ and stable if $E_{per} < E_{fluid}$.  The existence of the Wigner crystal corresponds to the statement that there is a range of low densities for which $E_{per}$ is smaller.  

This analysis also proves that the fluid crystal transition is first order.  The fluid solution a exists for some range of densities and is locally stable at large $N$.  It crosses over to a lower lying crystalline state at densities of order $1/N$ if the energy densities cross.  We do not have rigorous arguments in either direction about the existence of this crossing, though Wigner's original physical intuition would lead one to imagine that they do cross.  The failure of the leading order large $N$ approximation to see the crystal is entirely due to the fact that it treats the electron density as a continuous function which can go to zero smoothly.  

Another intriguing feature of the next to leading equations is the possibility that for $G_{per}^{(2)}$ develops negative eigenvalues in the periodic sector.  This must happen continuously as a function of density, which means that there is a critical density at
which we see a gapless mode in the correlator of the H-S field.  A strong argument that this indeed happens is that we expect that at some value of the density, probably of order $1$ in powers of $N$, the Hartree approximation becomes valid and the system becomes oblivious to the discreteness of the electron charge.  The meta-stable crystalline phase should cease to exist.  

This leads to several interesting conjectures.  Kivelson and Spivak\cite{ks} argued that in two dimensions the homogeneous electron fluid had at least one colloidal/microemulsion phase separating the crystalline and fluid phases.  A dimension independent variational argument for this was given in\cite{bz3}.  The authors of\cite{ks} also conjectured that the transition between the colloidal and fluid phases was second order.  Additional evidence for a mysterious second order transition out of the fluid phase was found in\cite{haule}\cite{tanaka}.    Given the existence of a meta-stable periodic solution of the next to leading order large $N$ SD equations in some range of densities, one is also guaranteed\cite{bz3} the existence of "critical bubble" solutions consisting of a marginally stable bubble of fluid trapped inside the crystal.  One would then hope to be able to construct a much more explicit version of the arguments of\cite{bz3} for the existence of the colloidal phase.  

Finally, the continuous disappearance of the negative eigenvalues of $G^{(2)}$ at a density above the liquid/crystal transition, signals the existence of a gapless excitation in the two point function of the density operator.  This might lead us to a systematic large $N$ treatment of the second order transition between fluid and colloidal phases of the homogeneous electron gas.  

We remind the reader of the very general argument for the existence of a colloidal phase that was given in\cite{bz3}.  First examine the critical bubble just above the density where the homogeneous fluid has lower energy density than the crystal.  It is large and its dynamics should be treatable by the macroscopic methods of elasticity theory.  Bubble expansion is resisted by elastic forces in the crystal, which for a single bubble are released by sending sound waves out to infinity.  However, for the multibubble configurations whose expansion and collision in a fluid, quickly converts one phase into another, we will instead get configurations with multiple large bubbles of fluid, of irregular shape, with thin crystalline walls separating them.  These will have energies that differ from the homogeneous fluid by terms that scale like a lower power of the volume.  

The Hamiltonian will have matrix elements between these different configurations.  Physically, this leads to movement of the bubbles over each other, turning the material into a {\it gel}.  Mathematically, it lowers the energy of the stable superposition of gel configurations.  Random matrix estimates suggest that it lowers it below the energy of the pure fluid.  As the density is increased further, the fraction of crystalline walls decreases, and a non-thermodynamic metal-insulator transition occurs (if the fluid is charged).  At this point the fate of the system depends on the surface tension of a configuration of crystal immersed in the fluid.  Kivelson and Spivak have argued\cite{ks} that it is negative in two dimensions, which implies that the crystal impurities are stable excitations.  In\cite{bz3} we conjectured that the same was true in three dimensions.  Overhauser's\cite{overh} Hartree-Fock calculation of a preferred charge density wave ground state provides weak evidence for this conjecture.  Whenever the surface tension is negative, its continuous transition to zero provides a mechanism for a second order transition from a colloidal to a homogeneous state.  We conjecture that the appearance of zero eigenvalues in $G^(2)$ is the signal of this transition in leading order in $1/N$.  

\section{Conclusions}

We have presented a general strategy for using Schwinger-Dyson equations to provide non-perturbative resummations of semi-classical and vector large N expansions.  In ordinary perturbation theory, the use of SD equations is not new, most notably in the Eliashberg equations of superconductivity theory, variations of which have also been applied in other condensed matter contexts\cite{electrons}.  Our methods are both more general and more systematic.  In principle they define an infinite hierarchy of non-perturbative resummations, each of which reduces to a finite number of integral equations.  Powerful numerical techniques can be brought to bear to solve the integral equations, although we have not attempted to do so in this paper.  In the case of standard phase transitions in scalar field theory, these equations provide a sequence of non-perturbative resummations of the $\epsilon$ expansion.  

The application of these ideas to large $N$ expansions is new, and presents a significant advance over the method proposed in\cite{tblargeNresum}, which required one to solve an integral equation for each value of the 't Hooft coupling and integrate the result w.r.t. the coupling, using the known weak coupling behavior as a boundary condition.  Although the method is applicable to the correlation functions of the Hubbard-Stratonovich field in any large $N$ vector model, we presented it explicitly for two dimensional models of critical bosons coupled to a fermi surface, and for the homogeneous electron gas.  The original hope was that it would find the correct IR behavior of Hertz-Millis models, but in the course of writing this paper we discovered an argument that suggests this is wrong. Our approximations systematically neglect higher $k$ point vertices of the single fermion loop graphs contributing to the effective action of the H-S field.  These vertices are small at large $N$ and fixed momenta, but dominate at small frequency and momenta, so they cannot be neglected in the IR regime.  

For the homogeneous electron gas the results are more promising.  The equations for the large $N$ saddle point have only a homogeneous solution and show no sign of a Wigner crystal phase. The energy of the classical saddle is exactly zero.  The first non-trivial correction to the SD hierarchy has both homogeneous and periodic solutions for the electron density.  It is reasonable to conjecture that the energies of the corresponding states, which are both of order $1$, cross at some point, corresponding to the first order transition to the Wigner crystal state.  If that is the case, these equations can be used to construct explicit solutions corresponding to the semi-classical colloidal configurations whose existence was argued for in\cite{bz3}.  Perhaps one can even come up with a more rigorous version of the argument\cite{bz3} that a superposition of these semi-classical states has lower energy than the homogeneous fluid, in a range of densities just above the first order transition.  Finally we identified an explicit mechanism by which a massless mode might appear in the correlator of the Hubbard-Stratonovich field in the meta-stable periodic phase.  If this actually occurs, we would have explicit tools to study the second order phase transition between colloidal and fluid phases conjectured in\cite{ks}\cite{haule}\cite{tanaka}.  
\vskip.3in
\begin{center}
{\bf Acknowledgments }
\end{center}
 The work of T.B. was supported by the Department of Energy under grant DE-SC0010008. Rutgers Project 833012.



\begin{thebibliography}{99}
 
\bibitem{sslee} S.S. Lee, "Low Energy Effective Theory of Fermi Surface Coupled with U(1) Gauge Field in 2 + 1 Dimensions" Phys. Rev. B80, 165102, (2009).
\bibitem{tblargeNresum} T.~Banks,``Systematic Resummation of the Large N expansion of Vector Models: Application to the Hubbard model and $2 + 1 $ dimensional QED,''
[arXiv:2003.13136 [hep-th]].
\bibitem{colloids} B Spivak, SA Kivelson, Phases intermediate between a two-dimensional electron liquid and Wigner crystal. Phys.Rev.B70.155114(2004); S.A. Kivelson, B. Spivak, Transport in two dimensional electronic micro-emulsions Ann. Phys. 321.9,2071-2115(2006); Jamei, Reza, S.Kivelson, Spivak, Boris, Universal Aspects of Coulomb-Frustrated Phase Separation. Phys. Rev. Lett. 94.5,056805(2005); T.~Banks and B.~Zhang,``Instantons, Colloids and Convergence of the 1/N Expansion for the Homogeneous Electron Gas,''
Annals Phys. \textbf{407}, 212-227 (2019)
doi:10.1016/j.aop.2019.05.003
[arXiv:1904.00251 [cond-mat.str-el]]; T.~Banks and B.~Zhang,
``On the low density regime of homogeneous electron gas,''
Annals Phys. \textbf{412}, 168019 (2020)
doi:10.1016/j.aop.2019.168019
[arXiv:1909.00291 [cond-mat.str-el]]; T.~Banks and B.~Zhang,``On the Colloidal Phase of the Homogeneous Electron Fluid,''
[arXiv:2010.05367 [cond-mat.str-el]].
\bibitem{ks} B Spivak, SA Kivelson, Phases intermediate between a two-dimensional electron liquid and Wigner crystal. Phys.Rev.B70.155114(2004); S.A. Kivelson, B. Spivak, Transport in two dimensional electronic micro-emulsions Ann. Phys. 321.9,2071-2115(2006); Jamei, Reza, S.Kivelson, Spivak, Boris, Universal Aspects of Coulomb-Frustrated Phase Separation. Phys. Rev. Lett. 94.5,056805(2005).
\bibitem{bz3} T.~Banks and B.~Zhang,``On the Colloidal Phase of the Homogeneous Electron Fluid,''
[arXiv:2010.05367 [cond-mat.str-el]].
\bibitem{bz1} T.~Banks and B.~Zhang, ``Instantons, Colloids and Convergence of the 1/N Expansion for the Homogeneous Electron Gas,''
Annals Phys. \textbf{407}, 212-227 (2019)
doi:10.1016/j.aop.2019.05.003
[arXiv:1904.00251 [cond-mat.str-el]].
\bibitem{overh} A.W.Overhauser, Spin density waves in an electron gas. Phys. Rev. 128.3,1437 (1962).
\bibitem{electrons} R.M. Martin, L. Reining, D.M. Ceperly, Interacting Electrons: Theory and Computational Approaches, (particularly Part IV), Cambridge University Press, (2016).
\bibitem{haule} K. Chen, K. Haule, Feynman's Solution of the Quintessential Problem in Condensed Matter Physics, arXiv:1809.04651.
\bibitem{tanaka}  Y. Takada, Excitonic Collective Mode and Negative Com- pressibility in Electron Liquids, J.Superconductivity,18.5- 6,785-789(2005).
 \end{thebibliography}


\end{document}